\documentclass[aps,prd,10pt,nofootinbib,twocolumn,eqsecnum,showpacs,showkeys,superscriptaddress,preprintnumbers,altaffilletter]{revtex4-1}

\usepackage{graphicx}
\usepackage{amsmath}
\usepackage{multirow}
\usepackage{hyperref}
\usepackage{soul}
\usepackage{graphicx}
\usepackage{dcolumn}
\usepackage{amssymb}
\usepackage{amsfonts}
\usepackage{amsbsy}
\usepackage{color}
\usepackage{rotating}
\usepackage[english]{babel}
\usepackage{multirow}
\usepackage{float}
\usepackage{orcidlink}

\usepackage[normalem]{ulem}

\hypersetup{
    colorlinks=true,
    linkcolor=blue,
    filecolor=magenta,
    urlcolor=blue,
    citecolor=blue
}

\newcommand{\be}{\begin{equation}}
\newcommand{\ee}{\end{equation}}
\newcommand{\bea}{\begin{eqnarray}}
\newcommand{\eea}{\end{eqnarray}}
\newcommand{\beal}{\begin{aligned}}
\newcommand{\eeal}{\end{aligned}}
\newcommand{\bi}{\begin{itemize}}
\newcommand{\ei}{\end{itemize}}

\defcitealias{Zamani:2024oep}{Paper~I}

\begin{document}

\title{Nonminimally coupled Dark Matter in Clusters of Galaxies: a fully comprehensive analysis}

\author{Saboura Zamani\,\orcidlink{0009-0004-3201-9483}}
\email{saboura.zamani@phd.usz.edu.pl}
\affiliation{Institute of Physics, University of Szczecin, Wielkopolska 15, 70-451 Szczecin, Poland}
\author{Vincenzo Salzano\, \orcidlink{0000-0002-4905-1541}}
\email{vincenzo.salzano@usz.edu.pl}
\affiliation{Institute of Physics, University of Szczecin, Wielkopolska 15, 70-451 Szczecin, Poland}
\author{Dario Bettoni\, \orcidlink{0000-0002-0176-5537}}
\email{dbet@unileon.es}
\affiliation{Departamento de Matemáticas, Universidad de León,
Escuela de Ingenierías Industrial, Informática y Aeroespacial
Campus de Vegazana, s/n
24071 León}
\affiliation{IUFFyM, Universidad de Salamanca, E-37008 Salamanca, Spain}

\date{Received: date / Accepted: date}

\begin{abstract}
In this study, we explore how a non-minimal coupling between dark matter and gravity can affect the behavior of dark matter in galaxy clusters. We have considered the case of a disformal coupling, which leads to a modification of the Poisson equation.
Building on an earlier work, we expand the analysis considering all possible disformal coupling scenarios and employing various dark matter density profiles. In doing so, we aim to constrain the key parameter in our model, the characteristic coupling length.
To achieve this, we analyze data from a combination of strong and weak lensing using three statistical approaches: a single cluster fitting procedure, a joint analysis, and one with stacked profiles. Our findings show that the coupling length is typically very small, thus being fully consistent with general relativity, although with an upper limit at $1\sigma$ which is of the order of $100$ kpc. 
\end{abstract}

\maketitle

\section{Introduction}
\label{sec: Intro}
It has been around ninety years since the indirect detection of ``Dark Matter'' (DM) by Fritz Zwicky \cite{Cirelli:2024ssz}. 
By studying the velocity dispersion of galaxies within the Coma cluster, he found that the observed mass was far less than what was needed to hold the cluster together
\cite{Zwicky:1933gu,Zwicky:1937zza}. As a result, he suggested the existence of an invisible form of matter that interacts with baryonic matter only through gravity.
In the subsequent years, more evidence supporting the existence of additional matter in the universe increased, such as rotational curves in galaxies discovered by Vera Rubin \cite{Rubin:1970zza, Rubin:1978kmz, Rubin:1980zd}, the gravitational lensing \cite{Brainerd:1995da,Paraficz:2012tv,Clowe:2006eq}, as well as Cosmic Microwave Background (CMB) anisotropies \cite{Planck:2018vyg}.

Since the beginning of the 21st century, the $\Lambda$CDM model, became widely recognized as the leading cosmological theory.
Based on GR, it has proven remarkably successful in various aspects. However, some challenges and unresolved issues remain \cite{Perivolaropoulos:2021jda, Abdalla:2022yfr, Akrami:2018vks, eBOSS:2020yzd, DES:2021esc, DES:2021es, Bull:2015stz, DiValentino:2021izs}, both at cosmological and astrophysical scales, the latter ones being our main focus in this work.

Considering GR as a special case, extended theories of gravity (ETGs) may offer a new framework that could potentially expand or replace the standard $\Lambda$CDM model and deepen our understanding of DM and dark energy (DE).  A diverse collection of models has emerged over time within the ETG category, each uniquely contributing to our understanding of these phenomena \cite{Nojiri:2017ncd, Clifton:2011jh, Koyama:2015vza, Ishak:2018his, Capozziello:2011et}.

Among them, and more specifically related to the DM problem, a well-known alternative to a DM particle solution is the MOND (Modified Newtonian Dynamics) paradigm \cite{Milgrom:1983a, Bekenstein:2010pt}, which seeks to explain this extra matter without including any DM. Although MOND was successful in some aspects \cite{Milgrom:2018fzq,Milgrom:2018tqf,Kelleher:2024oip}, it faces several challenges. Consequently, even within the MOND framework, some form of DM is still needed to fully reproduce observations across different scales \cite{Richtler:2007jd}.

In order to combine the positive aspects of the MOND approach with the strengths of the CDM model, a new model that brings together both ideas has been proposed in \cite{Bruneton:2008fk, Bettoni:2011fs, Bettoni:2015wla}.
There, the authors consider the concept of an effective non-minimal coupling between DM and gravity, which extends beyond the conventional interactions described in GR. Recently, a new study has appeared, exploring ultra-compact objects within the framework of this coupling \cite{Benetti:2024efb}.

This coupling between DM and gravity,  modifies both fluid dynamics and gravitational behavior and it should  be considered as arising from a ``coarse-graining" procedure, rather than originating from a fundamental modification of gravitational dynamics. Consequently, it doesn't introduce any new degrees of freedom.\footnote{Notice, however, that the equations of motion derived from this model are not higher order.}

At large cosmological scales, the NMC model is usually assumed to become  activated at late times, typically around $z \approx 5 $ \cite{Bettoni:2012}, so to reproduce the $\Lambda$CDM dynamics at earlier times. Hence, it is when considering smaller astrophysical scales, such as galaxy clusters and galaxies, that it might offer a deeper understanding of DM and may address some of the challenges that our standard cosmology model has at these scales.  Some analyses of the NMC model have been conducted at the level of galaxies, as discussed in \cite{Gandolfi:2022puw} and of galaxy clusters, where a study has been performed focusing only on X-ray observations \cite{Gandolfi:2023hwx}.

In this work we want to continue the analysis started in \citetalias{Zamani:2024oep}, aimed to explore the characteristics of DM seen as a fluid that is non-minimally coupled with gravity, as developed in \cite{Bettoni:2015wla}. We are particularly interested in results when moving to the Newtonian regime where the NMC produces a modified Poisson equation via an additional term of the form $\epsilon L^2 \nabla^2 \rho_{\mathrm{DM}}$. Herein $L$ represents the (dimensionful) coupling or characteristic length that measures the strength of the interaction. This modified equation indicates that both the density of DM and its distribution source the gravitational potentials.

An aspect it is worth emphasizing is that the Newtonian limit modifications obtained in the NMC scenarios are not limited a scenario in which DM behaves as a Bose-Einstein condensate \cite{Bettoni:2013zma,Ivanov:2019iec}, but can also be found in ETGs models, such as Born-Infeld theories, where similar modifications can be observed \cite{BeltranJimenez:2017doy}.

Following up with the analysis we started in \citetalias{Zamani:2024oep}, we aim at extending the study to disformal coupling with polarity $\epsilon=\pm1$, adding more DM mass models and an updated statistical analysis. The main motivation for that, spur from the most interesting result of \citetalias{Zamani:2024oep}, in which the coupling length of the disformal $\epsilon=-1$ scenario was very close to the NFW profile's characteristic radius, namely, $L \propto r_s$. There, we missed to understand if there was a deep physical reason after this result, or if it was just a (very) finely tuned statistical fluke. In light of what we are going to discuss in the following paragraphs and the more complete analysis we have performed, we have also re-analysed this case, in order to provide a global and self-consistent picture of the model. 

The paper is structured as follows. First, in Sec.~\ref{sec: Theo}, we present an overview of the theoretical background of the NMC model. In addition, since gravitational lensing serves as our tool for investigating the characteristics of DM, we review the foundational principles of gravitational lensing theory. Subsequently, we introduce the necessary modifications required by our model. In Sec.~\ref{sec: Matter}, we present the masses considered in our study. This section provides a brief discussion of the main density profile selected for our analysis and an overview of the gas density considered in our work. In Sec.~\ref{sec: SA}, we begin with a brief introduction to the dataset used in our analysis, namely the CLASH program. We then provide an overview of the statistical tools employed in this study. Finally, in Sections~\ref{sec: Discu} and \ref{sec: Conc}, we present the results, provide a comprehensive discussion, and present our conclusions. 
Given that our study includes multiple density profiles, \ref{sec: appendix} provides an overview of each.

%%%%%%%%%%%%%%%%%%%%%%%%%%%%
\section{Theory}
\label{sec: Theo}
In this section, we will briefly review the theoretical foundation of the NMC model we are considering in this work. The general action that characterizes the NMC model can be written as \cite{Bettoni:2015wla} 
\begin{equation}\label{eq: action_full}
S = S_{\rm EH} + S_{d} + S_{c} + S_{m}\, ,
\end{equation}
where $S_{\rm EH}$ stands for the Einstein--Hilbert action, $S_d$ reads 
\be
\label{eq: action}
        S_{d} = \frac{M_{\textrm{Pl}}^2}{2}\int d^4 x\sqrt{-g}
        \big[\alpha_{d} F_d (\rho) R_{\mu\nu}u^\mu u^\nu \big]\, ,
\ee
with the term $\alpha_d F_d (\rho) R_{\mu\nu} u^\mu u^\nu$ representing a disformal coupling between DM and curvature\footnote{We won't enter into the detailed exploration of the mechanisms behind disformal coupling. For our purposes, it suffices to know that these couplings can be generated via disformal transformations \cite{Bekenstein:1992pj} of the Einstein--Hilbert action. For a more comprehensive discussion on these topics, we refer to \cite{Bettoni:2011fs,Bettoni:2015wla}.}, 
and $S_{c}$ is
\be
\label{eq: action_conf}
        S_{c} = \frac{M_{\textrm{Pl}}^2}{2}\int d^4 x\sqrt{-g}
        \big[\alpha_{\mathrm{c}} F_c (\rho) R\big]\, ,
\ee
where the term $\alpha_c F_c (\rho) R$ represents a conformal coupling. Finally, $S_m$ encapsulates the dynamics of DM, which is derived from the perfect fluid action \cite{Brown:1992kc}, whose details are irrelevant to our discussion. 

The parameters $\alpha_d$ and $\alpha_c$ measure the strength of the non-minimal couplings while the functions $F_d(\rho)$ and $F_c(\rho)$ are, in general,  arbitrary function of the fluid variables. However, in this study, we adopt the simplest approach from \cite{Bettoni:2015wla} and define them $ F_i (\rho)\propto \rho_{{\rm DM}}$, thus excluding baryons which will behave as in the standard model.

Since in this work we are interested in the Newtonian limit, we will now briefly review the main modifications to the standard case introduced by the NMCs. Starting from the metric perturbations in the Newton gauge written as
\begin{equation}
    ds^2 = -(1+2 {\rm \Psi})dt^2+(1-2 {\rm \Phi})\delta_{ij}dx^idx^j\, ,
\end{equation}
we obtain the modified Poisson equation \cite{Bettoni:2011fs,Bettoni:2015wla}, which, in the case of a disformal coupling, can be written as
\be
\label{eq: poissonmoddisf}
    \mathbf{\nabla}^2 {\rm \Phi} = 4\pi G\,\Big[\rho_{\mathrm{tot}} - \epsilon\,\frac{L^2}{{2}}\, \nabla^2 \rho_{\rm DM}(r)  \Big]\, ,
\ee
where we have made the following identification $\alpha_d F_d = -8\pi G L^2\rho_{\rm DM}$. Furthermore, $\rho_{\mathrm{tot}} = \rho_{\rm bar} + \rho_{\rm DM}$, the total mass density of both baryonic matter and DM respectively, and the parameter $\epsilon = \pm 1$ serves as a dimensionless constant, representing the polarity of the coupling. 

It is worth to note that, in the case of a disformal coupling, we have ${\rm \Phi} = {\rm \Psi}$, namely, a disformally coupled DM fluid
can be described by a single gravitational potential and has no anisotropic stresses. The reason is that the disformal spatial coupling part is proportional to the spatial velocity $u^i$ and its derivatives, which are negligible in this limit (thus leading to no anisotropic stress). Thus the Laplacian of the Weyl potential, which is the quantity that can be effectively tested by using gravitational lensing, reads, in this case, 
\begin{equation}\label{eq: Weyl_disf}
\nabla^{2} \left( \frac{\mathrm{\Phi}+\mathrm{\Psi}}{2} \right) = 4\pi G\,\Big[\rho_{\rm tot} - \epsilon\,\frac{L^2}{{2}}\, \nabla^2 \rho_{\rm DM}(r)  \Big]\, .
\end{equation}
For the conformal coupling we arrive at a similarly modified Poisson equation:
\be
    \label{eq:poissonmodconf}
    \mathbf{\nabla}^2 {\rm \Phi} = 4\pi G\,\Big[\rho_{\rm tot} - \epsilon\,L^2\, \nabla^2 \rho_{\rm DM}(r)  \Big]\, ,
\ee
after the following identification $\alpha_c F_c = -8\pi G L^2\rho_{\rm DM}$. However, in this case, anisotropic stress is present, i.e., ${\rm \Phi} \neq {\rm \Psi}$, and the two gravitational potentials are related by
\begin{equation}\label{eq: Weyl_con}
{\rm \Psi} = {\rm \Phi} - \alpha_{c} F_c= \mathrm{\Phi}+8\pi G L^2\rho_{\rm DM}\, . 
\end{equation}
As a consequence, when calculating the Weyl potential, the NMC contributions cancel out, resulting in an equivalent GR equation. Thus, we will not explore the case of a conformal coupling in this study, because gravitational lensing is basically unaffected by this type of coupling \cite{Bettoni:2011fs}. 

Another interesting possibility is to consider a coupling to the Einstein tensor, that can be easily obtained by taking $\alpha_c = -\alpha_d/2$ and identifying $F_{c} = F_{d} = F_{E}$ in Eq.~(\ref{eq: action_full}). In this case the gravitational and interaction parts of the action read \cite{Bettoni:2011fs,Bettoni:2015wla}
\be
\label{eq: action_con}
        S = \frac{M_{\textrm{Pl}}^2}{2}\int d^4 x\sqrt{-g} \,\left[ R +\alpha_{E} F_{E}(\rho) G_{\mu\nu} u^\mu u^\nu \right]
        \; .
\ee
The interest in this particular scenario lies in the fact that in the Newtonian limit there are no modifications to the Poisson equation, as can be easily checked by combining equations \eqref{eq: poissonmoddisf} and \eqref{eq:poissonmodconf}. Hence, we are left with the same equation as in GR, but the anisotropic stress which is present in turns leads to a modified Weyl potential \cite{Bettoni:2015wla}. With this coupling, from Eq.~(\ref{eq: Weyl_disf}) and Eq.~(\ref{eq: Weyl_con}), it can be seen that the Weyl potential remains basically the same as in the disformal case. This means that the overall effect on gravitational lensing observations is similar to disformal coupling, even though the anisotropic stress differs, and then the two scenarios cannot be distinguished when using such kind of probe.

As can be easily seen, Eq. \eqref{eq: poissonmoddisf} shows that not only the total matter is the source of gravity, but also spatial inhomogeneities in DM distribution play a role. Consequently, the impact of this modification becomes more pronounced as the distribution of dark matter becomes more inhomogeneous \cite{Bettoni:2011fs}.

The quantity $L$ is considered as the characteristic length of the NMC model that is introduced for dimensional consistency \cite{Gandolfi:2021jai}. A priori, it does not need to be a universal constant, as it could depend on the characteristics of the local environment. But it also makes perfect sense to think about it as a universal length. In fact, in the case of Bose-Einstein condensates, it can be related to the microscopic properties of the boson(s), like mass and interactions, usually being referred to as healing length (see \cite{Bettoni:2013zma,Boehmer:2007um}). 

\subsection{Gravitational Lensing}
\label{sec: Lensing}

Galaxy clusters possess the great property to function as gravitational lenses, offering a chance to examine the structure of DM halos \cite{Kneib:2011jaf}. Also, and most importantly, DM is known to be the dominant component within galaxy clusters \cite{Diaferio:2008jh, Gonzalez:2013awy}. The remaining fraction consists of baryonic matter, primarily in the form of intracluster gas. This characteristic makes galaxy clusters highly suitable candidates for studying the properties of dark matter. 

The main theoretical details of gravitational lensing observables are quite well known  \cite{gralen.boo,Meneghetti2021LNPbook,Narayan:1996ba} and a summary can be also found in \citetalias{Zamani:2024oep}, so we only point out here the main equation, specifically, the expression of the convergence,
\be \label{eq: conv_GR}
    \kappa(R) = \frac{1}{c^2}\frac{D_{ls}D_l}{D_s}\int^{+\infty}_{-\infty}\nabla^2_r {\rm \Phi}(R,z)dz \, ,
\ee
which is the quantity that is actually reconstructed from observations. Here $R$ is the two-dimensional projected radius on the lens plane, and $r = \sqrt{R^2+z^2}$ represents the three-dimensional radius. Furthermore, $\nabla^2_r = \frac{2}{r}\frac{\partial}{\partial r} + \frac{\partial^2}{\partial r^2}$ stands for the radial part of the Laplacian in spherical coordinates, the only contributing term as we will assume spherical symmetry. The distances $D_{l}$, $D_{s}$ and $D_{ls}$ are the angular diameter distances between the observer and the lens, the observer and the source, and between the lens and the source, respectively. When needed to calculate them, we adopt the background cosmological parameters provided by the \textit{Planck} baseline model \cite{Planck:2018vyg} with Hubble constant $H_0 = 67.89$ km s$^{-1}$ Mpc$^{-1}$ and matter density parameter $\Omega_{m} = 0.308$.

By considering the Poisson equation, which in GR is written as
\begin{equation}
    \nabla^{2}_r {\rm\Phi} = 4\pi G_N\rho(r)\; ,
\end{equation}
one can establish a link between the convergence $\kappa$ and the total density $\rho$ within the lens system, and we end up with
\begin{equation}
    \label{eq: kappa}
    \kappa (R) = \int_{-\infty}^{+\infty} 
    \frac{4 \pi G_N}{c^{2}} \frac{D_{ls}D_{l}}{D_{s}}
    \rho(R,z)dz \equiv \frac{\Sigma(R)}{\Sigma_{cr}}\, ,
\end{equation}
where $\Sigma(R)$ is the two-dimensional surface mass density of the lens defined as 
\be
    \Sigma(R) = \int_{-\infty}^{+\infty} \rho(R,z) \mathrm{d}z\, ,
\ee
and $\Sigma_{cr}$ is the critical surface density of a gravitational lensing system
\be
   \Sigma_{cr} \,= \frac{c^{2}}{4\pi G_N} \, \frac{D_{s}}{D_{ls}D_{l}}\, .
\ee
Thus far, we have focused on GR, which does not have anisotropic stress, so that $ \rm \Phi = \Psi$. Yet, when dealing with ETGs, one should work with the more general version of the Weyl potential because it might be that $\rm \Phi \neq \Psi$. In this case, 
Eq. \eqref{eq: conv_GR} would be generalized as
\begin{equation}
 \label{eq: kappamod_1}
 \kappa(R) = \frac{1}{c^2}\frac{D_{ls}D_l}{D_s}\int^{+\infty}_{-\infty}\nabla^2_r\biggl\{\frac{{\rm \Phi}(R,z) + {\rm \Psi}(R,z)}{2}\biggr\}dz\, .
\end{equation}

Although for the disformal NMC model we consider here the potentials are the same, we do have a modified Poisson equation, so that Eq.~(\ref{eq: kappa}) now reads 
    \be
\label{eq: kappamod_Dis}
   \kappa (R) = \frac{1}{\Sigma_{cr}} \int_{-\infty}^{+\infty} 
    \left[\rho(R,z) - \epsilon \frac{L^2}{2} \nabla^2_r \rho_{\rm DM}(R,z)\right] \, \mathrm{d}z 
     \, .
     \ee

As can be seen, the convergence is affected by the Laplacian of the DM distribution, $\nabla^2_r \rho_{\rm DM}(R, z)$, which we can interpret as the lensing convergence being sensitive to how concentrated DM is at a given point.

The total density, denoted as $\rho$, comprises two components: $\rho_{\rm DM}$ representing the DM density and $\rho_{gas}$ representing the density of the hot intracluster gas. In the following sections, we will discuss more about the DM models we have chosen, and the reason for our choice of gas density.

\section{Mass Components}
\label{sec: Matter}

Galaxy clusters are the largest virialized structures in the sky. Comprising three main components, clusters consist of collisionless DM (dominant), hot diffuse baryons (X-rays), and cooled baryons, such as stars within galaxies \cite{Voit:2004ah}. Throughout this work, we have modeled only DM and gas for their most relevant role in the mass budget.

\subsection{DM halo density Profiles}
\label{sec: DM}

One of the mostly used DM halo density profiles is the Navarro--Frenk--White (NFW) Profile, derived from N-body simulations using CDM cosmology \cite{NFW:1996}. The spherically symmetric NFW mass density profile is 
\begin{equation}
\label{eq: NFW}
    \rho_\mathrm{NFW}(r) =  \frac{\rho_{s}}{\frac{r}{r_{s}} \Big( 1 + \frac{r}{r_{s}} \Big)^{2}} \,,
\end{equation}
where $\rho_{s}$ denotes the characteristic density of the halo, and $r_{s}$ representing the scale radius. The scale density, $\rho_{s}$, can be expressed as follows
\begin{equation}
\label{eq: delta}
    \rho_{s} = \frac{\mathrm{\Delta}}{3} \rho_{c} \frac{c_{\mathrm{\Delta}}^{3}}{\mathrm{ln}(1+ c_{\mathrm{\Delta}}) - \frac{c_{\mathrm{\Delta}}}{1 + c_{\mathrm{\Delta}}}} \, ,
\end{equation}
where $ c_{\mathrm{\Delta}} = \frac{r_{\mathrm{\Delta}}}{r_{s}}$ is the dimensionless concentration parameter and $r_{\mathrm{\Delta}}$ represents the spherical radius at which the average density of the enclosed mass is $\mathrm{\Delta}$ times the critical density $\rho_c$ of the Universe at the lens redshift. Throughout this work and to perform a consistent comparison among the various DM profiles we use, we have normalized all the models to the same radius, $r_{-2}$, defined as the distance from the center where the logarithmic slope of the density profile becomes 
$
\frac{d \ln \rho(r)}{d \ln r} = -2\, .
$
Thus, the concentration parameter will be always defined as
\begin{equation}
c_{{\rm \Delta}} = \frac{r_{{\rm \Delta}}}{r_{-2}}.
\end{equation}
In the case of a NFW profile, we have $r_{s} = r_{-2}$, but this does not hold true for all the DM profiles we consider (see \ref{sec: appendix}). 

Additionally, $M_{\mathrm{\Delta}}$ denotes the total mass contained within the overdensity radius $r_{\mathrm{\Delta}}$, which for a NFW profile reads
\begin{equation}
\label{eq: M_delta}
    M_{\mathrm{\Delta}} = \frac{4}{3} \pi r_{\mathrm{\Delta}}^{3} \mathrm{\Delta} \rho_{c} = 4 \pi \rho_{s} r_{s}^{3} \bigg[ \mathrm{ln}(1+ c_{\mathrm{\Delta}}) - \frac{c_{\mathrm{\Delta}}}{1 + c_{\mathrm{\Delta}}} \bigg]\, .
\end{equation}
Throughout this work, we have considered the value of $\mathrm{\Delta} = 200$, which is the most standard choice in the literature. Consequently, the NFW parameters we have employed in our study are $\{c_{200}, M_{200}\}$.

Actually, we do not confine ourselves only to the NFW profile to describe DM, but we consider a variety of models. The main reason is that we want to understand if the result we obtained in \citetalias{Zamani:2024oep}, i.e., that $L\propto r_s$, was eventually due to a fine tuning \textit{also} related to the choice of the functional form for the density profile. Indeed, we have to stress that the NFW model emerges as a universal profile from GR-based simulation, so that it might not be the best model for the NMC scenario. In absence of detailed simulations based on this alternative model, we try to dig more in its behaviour by consider several DM profiles thus extending what done in \citetalias{Zamani:2024oep}. More specifically, we will consider: the Hernquist model \cite{Hernquist:1990be}; the Burkert model \cite{Burkert:2000di,Mori:2000me}; a cored NFW (cNFW) model \cite{Newman:2012nv,Newman:2012nw}; a generalized NFW (gNFW) model \cite{Umetsu:2015baa,Zhao:1995cp}; a DARK-exp-$\gamma$ model \cite{Umetsu:2015baa,Hjorth:2015bfa}; the Einasto profile \cite{Einasto:1965czb,Retana-Montenegro:2012dbd}; and a generalized pseudo-isothermal (GPI) one. For the sake of clarity and readability, details about each model are postponed to the \ref{sec: appendix}.

\subsection{Hot Gas}
\label{sec: Gas}
After launching the X-ray telescopes in early 1970, it was revealed that clusters serve as potent emitters of X-rays, with X-ray luminosities ranging between $10^{43} - 10^{45} \; \; \mathrm{ergs}^{-1}$. Given the assumption that the intra-cluster gas remains in hydrostatic equilibrium within a spherically symmetric gravitational potential encompassing all cluster matter, the X-ray temperature and flux serve as reliable indicators for estimating the mass of the cluster.

Although it is feasible to use X-ray observations for the clusters we consider, all of which have associated archival data \cite{Donahue:2014qda}, we have chosen not to incorporate them directly. The reason behind this decision arises from a well-known bias. It is acknowledged that X-ray data can be sensitive to non-gravitational local astrophysical phenomena, unlike gravitational lensing, which provides a pure gravitational probe. As a result, we have decided to prioritize a more reliable and less biased reconstruction approach, even if it entails sacrificing some precision.

Nevertheless, we consider hot gas in our cluster modeling, taking into account the gas density ($\rho_{gas}$) in the overall density presented in Eq. (\ref{eq: kappamod_Dis}). 
We adopt the approach outlined in \cite{Donahue:2014qda} using data from \textit{Chandra} and applying (at most) a truncated double $\beta$-model \cite{Cavaliere:78ab}
\begin{align}\label{eq: gas_dens}
\rho_\mathrm{gas}(r) &=  \rho_{e,0}\biggl(\frac{r}{r_0}\biggr)^{-\alpha}\biggl[1 + \biggl(\frac{r}{r_{e,0}}\biggr)^2\biggr]^{-3\beta_0/2} \nonumber \\
& + \rho_{e,1}\biggl[\biggl(\frac{r}{r_{e,1}}\biggr)^2\biggr]^{-3\beta_1/2}\, 
\end{align}
to fit the gas densities when a single $\beta$-model (or truncated $\beta$-model) does not fit properly the data. It should be considered that the free parameters in Eq.~(\ref{eq: gas_dens}), $\{\rho_{e,0},\rho_{e,1},r_0,r_{e,0},r_{e,1},\alpha,\beta_0,\beta_1\}$ are initially determined through independent fits and are not considered free parameters in our analysis.

\section{Statistical Analysis}
\label{sec: SA}

The data used in this paper are from the \textit{CLASH} (Cluster Lensing And Supernova survey with Hubble) program\footnote{\url{https://archive.stsci.edu/prepds/clash/.}} \cite{CLASH2012}. The clusters in the sample span the redshift range $0.18 < z < 0.90$, and the mass range $0.5 \lesssim M_{200}/(10^{15} M_\odot) \lesssim 3$. We have a total of $19$ clusters, among which 16 were identified using X-ray observations, whereas the other 3 were selected based on lensing data \cite{Umetsu:2015baa, Umetsu:2014vna, Merten:2014wna, Zitrin2015}.

In this sample, nearly half of the clusters are likely to be unrelaxed, like A209, RXJ2248, MACSJ1931, MACSJ0416, MACSJ1149, MACSJ0717, and MACSJ0647 \cite{Umetsu:2015baa}.
This means that the assumption of hydrostatic equilibrium, which is often used to estimate cluster masses, may not be fully valid for some of them \cite{Meneghetti:2014xna}. Although, in \cite{Umetsu:2015baa,Umetsu:2016cun}, it has been shown that the averaged surface mass density $\Sigma(R)$ and the convergence $\kappa(R)$ of the \textit{CLASH} X-ray-selected sample can be fairly modeled by the NFW profile within GR, so that the hydrostatic equilibrium can be pretty safely assumed.

\subsection{Single fits analysis}

As our goal is to constrain the parameters of the NMC model and those defining the DM profile for each individual cluster, we need to define the $\chi^2$ function as follows
\be
\label{eq: chi}
    \chi^{2}_{i}(\boldsymbol\theta_{i}) = {\rm \Delta} \boldsymbol{\kappa_{i}}(\boldsymbol\theta_{i}) \cdot \mathbf{C}^{-1}_{i} \cdot {\rm \Delta} \boldsymbol{\kappa_{i}}(\boldsymbol\theta_{i})\, ,
\ee
where: $\boldsymbol\theta_{i} = \{c_{200,i},\, M_{200,i}, \gamma_i, \, L_i\}$ represents the maximum set of free parameters (for each cluster, $i=1,\ldots,19$) when dealing with the NMC model and any of the chosen DM profiles (of course, within GR, $L_i=0$); ${\rm \Delta}  \boldsymbol{\kappa_i} = \boldsymbol{\kappa^{theo}_i}- \boldsymbol{\kappa^{obs}_i}$, is the difference between the theoretical and observed values of convergence; $\boldsymbol{\kappa^{obs}_i}$ and $\boldsymbol{\kappa^{theo}_i}$ both are vectors  made of 15 elements; $\boldsymbol{\kappa^{theo}_i}$ is calculated using Eq.~(\ref{eq: kappamod_1}) for GR, Eq.~(\ref{eq: kappamod_Dis}) for the disformal NMC case; finally, $\mathbf{C}_{i}$ is the covariance matrix \cite{Umetsu:2015baa,Umetsu:2020wlf}.

In order to minimize the $\chi^2$ function, we employed our custom Markov Chain Monte Carlo (MCMC) code. To ensure proper convergence of the chain, we followed the approach presented in \cite{Dunkley:2004sv}. For each cluster, we chose flat priors on the NFW parameters, ensuring that both the concentration parameter $c_{200}$ and the mass parameter $M_{200}$ are positive, reflecting physically meaningful values. We did not consider any priors for the characteristic length $L$, but we have chosen to work with $\log L$ as free parameter in our MCMCs, as we ignore the scale of this length, and the logarithmic scale allows us to sample consistently a large range of orders of magnitudes.

To evaluate the validity of our model in comparison to the standard theory (GR), we first computed the Bayesian evidence $\mathcal{E}$ using our customized implementation of the nested sampling algorithm described in \cite{Mukherjee:2005wg}.
We then calculated the Bayes factor $\mathcal{B}^{\,i}_{j}$ \cite{Kass:1995loi}, which is defined as the ratio of the Bayesian evidence for our NMC model $\mathcal{M}_i$ to that of the reference model $\mathcal{M}_j$ (GR).
In our analysis, we interpret the Bayes Factor using the empirical Jeffrey's scale \cite{JeffreysScale}. According to this scale:
$\ln \mathcal{B}^{i}_{j}<1$ means the evidence in favor of model $i$ is weak against model $j$;  $1<\ln \mathcal{B}^{i}_{j}<2.5$ the evidence is weak; $2.5<\ln \mathcal{B}^{i}_{j}<5$ evidence is moderate; and $\ln \mathcal{B}^{i}_{j}>5$ is strong.

\subsection{Joint analysis: universal \texorpdfstring{$L$}{L}}

In order to get even more insights into the problem, we have performed also two additional types of analysis. The first  and most straightforward generalization has considered the possibility to have only one single \textit{universal} interaction length $L$ for all clusters. In this case, the total $\chi^2$ will be defined as
\begin{equation}
\chi^2(\boldsymbol\theta) = \sum_{i=1}^{19} \chi^{2}_{i}(\boldsymbol\theta_i)\, ,
\end{equation}
where now the vector of parameters is made of (max.) $58$ elements, i.e., $\boldsymbol{\theta}=\{c_{200,i},M_{200,i}, \gamma_i, L\}$.

\subsection{Stacked profiles analysis}

Finally, once the direct analysis of the observed convergence profiles is performed, the constraining power of the data set has been enhanced through a stacking procedure. Such a method indeed washes out the possible astrophysical contamination of the lensing signal due to the peculiar physical state of the analyzed systems \cite{Okabe:2009pf}. A universal profile, with an emphasized signal-to-noise ratio and dependence on the cosmological background, is provided accordingly. The stacking technique is usually applied in the literature to galaxy cluster images to extract an improved signal-to-noise ratio in the outer part of the mass profiles \cite{Wilcox:2015kna,Sakstein:2016ggl}. The employment of this procedure is safe as long as the free parameters involved in the analysis only depend on the shape of the profiles and are not sensitive to their amplitude.

In this work, we apply the stacking procedure directly to the lensing convergence profiles instead of the raw data, as done in \cite{Umetsu:2010rv}. The key steps in our workflow are as follows:
\begin{enumerate}
    \item for each cluster:
    \begin{enumerate}
    \item we rescale the original radii base $R$ at which the convergence $\kappa(R)$ is measured ($N = 15$ elements) with respect to the spherical radius $r_{-2,i}$, getting $R \rightarrow R \equiv R/r_{-2,i}$. We have chosen to normalize the profiles of each cluster at their corresponding $r_{-2,i}$ obtained from a GR analysis assuming a NFW profile, but we emphasize here that this choice is totally arbitrary and with no impact on the final results;
    \item we find a new base of $M = 10$ radial bins chosen adaptively to be the best choice for all the clusters of the \textit{CLASH} sample and to match the original data distribution, where we compute the new stacked $\kappa(R)$ map;
    \end{enumerate}
    \item we construct the projection matrix $P_{ji}\in \mathbb{R}^{M\times N}$, which allows changing basis from the original dataset ($R_i$, $i=\{1,...,N\}$) to the newly defined one ($\tilde{R}_j$, $j=\{1,...,M\}$). The matrix elements are unambiguously determined via mass conservation, provided that the mass density is taken to be constant in each radial bin;
    \item each convergence profile is projected onto the common basis through the following relation:
    \be
    \Tilde{\kappa}_j = \mathcal{P}_{ji} \, \kappa_i \, ,
    \ee
    while the projected covariance matrix is given by
    \be
    \Tilde{C}_{kl} =\mathcal{P}_{ki} \, C_{ij} \, \mathcal{P}^T_{jl} \, ;
    \ee
    \item For each cluster, we calculate the projected mass density $\tilde{\Sigma}(\tilde{R})$ profile. We have decided to use $\Sigma(\tilde{R})$, instead of $\tilde{\kappa}(\tilde{R})$, because it depends mostly on the astrophysical properties of the system, and not on strongly-cosmological-dependent terms such as the critical density $\Sigma_c$. The stacked $\tilde{\Sigma}(\tilde{R})$ is defined as
    \begin{equation}
    \label{eqn: stack_sigma}
        \langle\mathbf{\tilde{\Sigma}}\rangle = \bigg( \sum_n \mathbf{\Tilde{C}}^{-1}_n \omega_n^{-2} \bigg)^{-1} \bigg( \sum_n \mathbf{\Tilde{C}}^{-1}_n \omega_n^{-1} \mathbf{\Tilde{\kappa}}_n \bigg)
    \end{equation}
    and the new stacked covariance matrix reads 
    \begin{equation}
    \label{eqn: covar}
        \mathbf{\tilde{\mathcal{C}}} = \bigg( \sum_n \mathbf{\Tilde{C}}^{-1}_n \omega_n^{-2} \bigg)^{-1}
    \end{equation}
    where $\omega_n = (\Sigma_c)_n$ are the weights with $n = 1,....19$ being the total number of clusters in the \textit{CLASH} sample;
    \item Eqs.~\eqref{eqn: stack_sigma} and \eqref{eqn: covar} represent the new data that will enter in the $\chi^2$,
    \be
\label{eq: chi_stack}
    \chi^{2}(\boldsymbol\theta) = {\rm \Delta}  \boldsymbol{\tilde{\Sigma}}(\boldsymbol\theta) \cdot \mathbf{\tilde{\mathcal{C}}}^{-1} \cdot {\rm \Delta} \boldsymbol{\tilde{\Sigma}}(\boldsymbol\theta)\, ,
\ee
    and that have to be compared with the theoretical definition of $\tilde{\Sigma}(\tilde{R})$ represented by the numerator of Eq.~\eqref{eq: kappa};
    \item the $\chi^2$ minimization has been performed using our MCMC algorithm and the constraints on the parameters $\boldsymbol{\theta} = \{M_{200}, c_{200}, \gamma, L\}$ for each DM profile. The conversion from dimensionless $R/r_{-2}$ to physical radii $R$ required to calculate the numerator of Eq.~\eqref{eq: kappa}, needs the definition of the angular diameter distance to a lens. For that, we have evaluated a mean redshift $ \langle z \rangle$ defined as the average of the redshift of each cluster weighted by the factor $w_n$.
\end{enumerate}

{\renewcommand{\tabcolsep}{1.mm}
{\renewcommand{\arraystretch}{2.}
\begin{table*}[h]
\begin{minipage}{\textwidth}
\centering
\caption{\textit{CLASH} clusters ordered by redshift. Our results regarding $c_{200}$, $M_{200}$ and $L$ are presented for both GR and modified Disformal case ($\mathrm{\Phi = \Psi}$), considering the combined contributions of DM and gas. Both single-fit and joint-fit analyses are conducted using the NFW density profile.}
\label{tab:results}
\scriptsize
\resizebox*{\textwidth}{!}{
\begin{tabular}{c|cc|ccc|ccc|ccc|ccc}
\hline
 & \multicolumn{8}{c|}{Single-Fit} & \multicolumn{6}{c}{Joint-Fit} \\
\hline
Cluster & \multicolumn{2}{c|}{GR} & \multicolumn{3}{c|}{$\epsilon = -1$} & \multicolumn{3}{c|}{$\epsilon = +1$} & \multicolumn{3}{c|}{$\epsilon = -1$} & \multicolumn{3}{c}{$\epsilon = +1$} \\
 & $c_{200}$ & $M_{200}$ 
 & $c_{200}$ & $M_{200}$ & $\log L$ 
 & $c_{200}$ & $M_{200}$ & $\log L$ 
 & $c_{200}$ & $M_{200}$ & $\log L$ 
 & $c_{200}$ & $M_{200}$ & $\log L$ \\
 &           & $(10^{15}\,\mathrm{M}_{\odot})$ 
 &           & $(10^{15}\,\mathrm{M}_{\odot})$ & (kpc)
 &           & $(10^{15}\,\mathrm{M}_{\odot})$ &  (kpc)
 &           & $(10^{15}\,\mathrm{M}_{\odot})$ & (kpc)
 &           & $(10^{15}\,\mathrm{M}_{\odot})$ & (kpc)
 \\
\hline
A383 & 
$6.52^{+2.75}_{-1.93}$ & $0.69^{+0.29}_{-0.22}$ & $6.57^{+2.53}_{-2.32}$ & $0.69^{+0.27}_{-0.22}$ & $<3.57$  &  $6.52^{+2.78}_{-1.95}$ & $0.70^{+0.29}_{-0.23}$ & $<3.23$ & $6.46^{+3.27}_{-1.94}$ & $0.70^{+0.26}_{-0.25}$& \multirow{19}{*}{$-168^{+130}_{-554}$} & $7.18^{+4.29}_{-3.01}$ & $0.86^{+0.36}_{-0.19}$ & \multirow{19}{*}{$3.21^{+0.07}_{-0.07}$} \\
A209 & 
$2.46^{+0.70}_{-0.57}$ & $1.58^{+0.48}_{-0.40}$ & $2.47^{+0.70}_{-0.59}$ & $1.57^{+0.47}_{-0.40}$ & $<3.03$  & $2.47^{+1.87}_{-0.21}$ & $1.58^{+0.39}_{-0.14}$ & $<3.70$  & $2.49^{+0.65}_{-0.58}$ & $1.59^{+0.50}_{-0.44}$ & 
& $2.40^{+0.82}_{-0.60}$ & $1.69^{+0.47}_{-0.45}$&  \\
A2261 &
$3.96^{+1.19}_{-0.94}$ & $1.97^{+0.57}_{-0.46}$ & $3.98^{+1.10}_{-0.97}$ & $1.97^{+0.57}_{-0.45}$ & $<3.56$  &  $3.95^{+1.14}_{-0.96}$ & $1.97^{+0.57}_{-0.45}$ & $<3.22$  & $4.03^{+1.18}_{-0.94}$ & $2.29^{+0.58}_{-0.52}$ &  & $4.36^{+1.72}_{-1.08}$ & $2.31^{+0.54}_{-0.57}$ & \\
RXJ2129 &
$6.64^{+2.63}_{-1.90}$ & $0.47^{+0.17}_{-0.14}$ & $6.60^{+2.37}_{-2.96}$ & $0.47^{+0.17}_{-0.13}$ & $<3.65$ & $6.61^{+2.47}_{-1.94}$ & $0.47^{+0.18}_{-0.14}$ & $<0.63$ & $5.96^{+2.28}_{-1.67}$ & $0.62^{+0.23}_{-0.18}$ & & $8.38^{+4.19}_{-3.07}$ & $0.73^{+0.16}_{-0.14}$ & \\
A611 &
$4.31^{+1.67}_{-1.26}$ & $1.37^{+0.51}_{-0.40}$ & $4.31^{+1.45}_{-1.16}$ & $1.36^{+0.45}_{-0.37}$ & $<0.54$ &  $4.30^{+1.61}_{-1.23}$ & $1.37^{+0.49}_{-0.40}$ & $<3.13$  & $4.29^{+1.81}_{-1.21}$ & $1.55^{+0.57}_{-0.45}$& & $5.09^{+3.91}_{-1.88}$ & $1.58^{+0.70}_{-0.42}$& \\
MS2137 &
$3.53^{+3.61}_{-1.69}$ & $0.93^{+0.68}_{-0.44}$ &  $3.48^{+2.57}_{-1.81}$ & $0.94^{+0.74}_{-0.39}$ & $<3.62$ &  $3.45^{+2.74}_{-1.81}$ & $0.94^{+0.78}_{-0.39}$ & $<3.00$  & $3.38^{+2.62}_{-1.46}$ & $1.16^{+0.80}_{-0.47}$ &  & $3.71^{+3.02}_{-1.72}$ & $1.20^{+0.64}_{-0.39}$ & \\
RXJ2248 & 
$4.63^{+2.45}_{-1.71}$ & $1.23^{+0.62}_{-0.42}$ & $4.56^{+2.17}_{-1.48}$ & $1.23^{+0.50}_{-0.41}$ & $<3.38$ &  $4.62^{+2.48}_{-1.72}$ & $1.22^{+0.59}_{-0.44}$ & $<3.04$ &  $4.56^{+2.33}_{-1.60}$ & $1.62^{+0.76}_{-0.53}$ & & $6.64^{+5.48}_{-2.90}$ & $1.39^{+0.67}_{-0.30}$ &  \\
MACSJ1115 &
$2.99^{+1.07}_{-0.79}$ & $1.44^{+0.45}_{-0.38}$ & $3.02^{+0.93}_{-0.89}$ & $1.44^{+0.47}_{-0.36}$ & $<-3.67$ &  $3.01^{+1.22}_{-0.85}$ & $1.44^{+0.45}_{-0.35}$ & $<3.58$  & $3.08^{+1.04}_{-0.78}$ & $1.78^{+0.53}_{-0.45}$ & & $3.27^{+1.60}_{-0.87}$& $1.78^{+0.49}_{-0.47}$&  \\
MACSJ1931 &
$4.97^{+3.74}_{-2.07}$ & $1.19^{+0.81}_{-0.52}$ & $4.99^{+3.63}_{-2.08}$ & $1.17^{+0.77}_{-0.51}$ & $<3.50$  & $4.99^{+3.73}_{-2.11}$ & $1.18^{+0.78}_{-0.52}$ & $<3.26$  & $4.42^{+2.98}_{-1.69}$ & $1.59^{+1.04}_{-0.65}$ & & $6.37^{+4.30}_{-2.60}$ & $1.41^{+0.61}_{-0.38}$ & \\
MACSJ1720 &
$5.08^{+2.04}_{-1.46}$ & $1.07^{+0.39}_{-0.32}$ & $5.08^{+1.90}_{-1.52}$ & $1.07^{+0.40}_{-0.31}$ & $<2.72$  &  $5.11^{+1.86}_{-1.55}$ & $1.07^{+0.40}_{-0.30}$ & $<3.23$  &  $4.82^{+1.75}_{-1.16}$ & $1.33^{+0.39}_{-0.38}$ & & $7
7.70^{+4.55}_{-3.05}$ & $1.23^{+0.36}_{-0.24}$ & \\
MACSJ0416 &
$3.12^{+0.92}_{-0.74}$ & $0.92^{+0.27}_{-0.23}$ &$3.13^{+0.88}_{-0.86}$ & $0.92^{+0.28}_{-0.23}$ & $<3.50$  &  $3.17^{+7.55}_{-3.96}$ & $0.93^{+0.40}_{-0.12}$ & $<3.46$  &  $3.05^{+0.80}_{-0.65}$ & $1.15^{+0.28}_{-0.27}$ & & $3.29^{+1.31}_{-0.79}$& $1.13^{+0.30}_{-0.26}$& \\
MACSJ0429 &
$5.79^{+2.84}_{-1.96}$ & $0.70^{+0.32}_{-0.23}$ & $5.82^{+2.48}_{-2.72}$ & $0.71^{+0.29}_{-0.22}$ & $<3.72$  &  $5.83^{+9.04}_{-1.45}$ & $0.70^{+0.25}_{-0.16}$ & $<3.33$  & $5.11^{+2.31}_{-1.64}$ & $0.95^{+0.42}_{-0.29}$ &  &$9.08^{+5.69}_{-4.49}$ & $0.99^{+0.29}_{-0.22}$ & \\
MACSJ1206 &
$4.74^{+2.06}_{-1.43}$ & $1.28^{+0.42}_{-0.35}$ & $4.77^{+1.62}_{-1.62}$ & $1.28^{+0.42}_{-0.30}$ & $<3.71$  &  $4.77^{+1.79}_{-1.58}$ & $1.28^{+0.46}_{-0.32}$ & $<3.46$  &  $4.23^{+1.48}_{-1.13}$ & $1.75^{+0.48}_{-0.44}$ &  & $5.35^{+4.24}_{-1.76}$ & $1.63^{+0.42}_{-0.33}$& \\
MACSJ0329 & 
$8.91^{+3.44}_{-2.41}$ & $0.64^{+0.18}_{-0.15}$ & $8.87^{+2.47}_{-5.54}$ & $0.64^{+0.18}_{-0.30}$ & $<3.84$  &  $8.82^{+7.44}_{-1.38}$ & $0.65^{+0.11}_{-0.05}$ & $<3.08$  & $7.73^{+2.84}_{-2.08}$ & $0.88^{+0.23}_{-0.21}$ &  & $10.02^{+3.85}_{-3.19}$ & $1.08^{+0.21}_{-0.18}$& \\
RXJ1347 &
$3.13^{+1.16}_{-0.85}$ & $2.96^{+0.94}_{-0.80}$ & $3.16^{+1.11}_{-0.92}$ & $2.95^{+0.98}_{-0.77}$ & $<-0.76$ &  $3.17^{+1.22}_{-0.89}$ & $2.93^{+0.95}_{-0.73}$ & $<3.66$  &  $3.36^{+1.21}_{-0.80}$ & $3.52^{+0.99}_{-0.89}$& & $3.58^{+1.42}_{-1.05}$& $3.46^{+1.17}_{-0.92}$& \\
MACSJ1149 &
$2.55^{+0.97}_{-0.71}$ & $1.81^{+0.56}_{-0.50}$ & $2.55^{+0.85}_{-0.81}$ & $1.80^{+0.63}_{-0.45}$ & $<0.33$ &  $2.55^{+1.17}_{-0.78}$ & $1.81^{+0.60}_{-0.44}$ & $<3.66$ &  $2.30^{+0.75}_{-0.60}$ & $2.54^{+0.61}_{-0.58}$ & & $2.45^{+0.85}_{-0.65}$ & $2.45^{+0.57}_{-0.53}$ & \\
MACSJ0717 &
$1.78^{+0.45}_{-0.38}$ & $2.55^{+0.61}_{-0.55}$ & $1.79^{+0.43}_{-0.39}$ & $2.53^{+0.64}_{-0.53}$ & $<3.61$ &  $6.06^{+3.12}_{-0.05}$ & $3.40^{+1.33}_{-0.27}$ & $3.61^{+0.12}_{-0.04}$ & $1.91^{+0.43}_{-0.38}$ & $2.90^{+0.65}_{-0.64}$ & & $1.94^{+0.51}_{-0.34}$ & $2.84^{+0.79}_{-0.58}$ & \\
MACSJ0647 &
$4.66^{+2.27}_{-1.56}$ & $1.20^{+0.46}_{-0.37}$ & $4.57^{+1.85}_{-1.67}$ & $1.21^{+0.49}_{-0.33}$ & $<3.72$ &  $4.57^{+1.40}_{-1.43}$ & $1.21^{+0.40}_{-0.27}$ & $<-1.30$ & $4.48^{+2.49}_{-1.39}$ & $1.45^{+0.52}_{-0.47}$ &  &$7.73^{+6.32}_{-3.58} $& $1.36^{+0.35}_{-0.31}$ & \\
MACSJ0744 &
$4.58^{+2.15}_{-1.38}$ & $1.31^{+0.46}_{-0.38}$ & $4.64^{+1.68}_{-1.65}$ & $1.31^{+0.53}_{-0.33}$ & $<1.11$ &  $4.61^{+3.37}_{-1.59}$ & $1.31^{+0.51}_{-0.32}$ & $<3.36$ &  $4.47^{+1.89}_{-1.41}$& $1.60^{+0.51}_{-0.43}$ & &$7.02^{+7.77}_{-2.98}$ &$1.54^{+0.33}_{-0.27}$ & \\
\hline
\end{tabular}}
\end{minipage}
\end{table*}}}

\section{Results and Discussion}
\label{sec: Discu}

Based also on literature, our reference model for comparison will be the NFW profile. All the results obtained from using NFW are presented in Table~\ref{tab:results}. In the following, we are going to discuss all the possible cases we have considered.

\subsection{Single fits with NFW: GR vs NMC scenario}

Upon looking at the left half of Table~\ref{tab:results}, where we show the results obtained assuming one characteristic $L$ per cluster, one can easily see that the values of $c_{200}$ and $M_{200}$, apparently, remain relatively consistent across all scenarios. This is equivalent to saying that 
NMC DM models are fully consistent with GR and barely deviate from it as confirmed also by the Bayes Factor (which we do not report in the table just for the sake of readability, but are available upon request) which is in all cases consistent with zero. This holds true also for the case with $\epsilon=-1$ previously considered in \citetalias{Zamani:2024oep}. 

The main difference here is that we have decided to take a much more conservative approach from the statistical point of view. First of all, we have run much longer MCMCs, at least doubling the number of points through which we sample the parameter space, passing from $\approx5 \cdot 10^4$ to $1-2 \cdot 10^5$ steps. Then, we have varied both the initial points of the MCMCs and the covariance matrix which dictates the size of the steps with which the MCMCs explore the space. After all such cases, we think that the conclusions we get are the most conservative and safe which we can get from the statistical point of view.

The challenging part when working with NMC DM still is, like in \citetalias{Zamani:2024oep}, the detection and the interpretation, in most cases, of a shift between the median value of $L$ and its value at the minimum of the $\chi^2$. As a result, the median is generally consistent with GR, returning values of $L$ which are typically very small (implying that the NMC correction to the Poisson equations are negligible), while the minimum \textit{may} lie much closer to larger values of $L\sim r_s$.

To address this point, in 
\citetalias{Zamani:2024oep}
we employed a Profile Distribution analysis \cite{Gomez-Valent:2022hkb, Trotta:2017wnx} in order to give much weight to the parameter space region around the minimum,  to explore it better and to eventually infer some statistical considerations. Unfortunately, with much larger MCMCs and after taking into account all the above checks, the Profile Distribution approach turns out to be not useful: the relative number of points which are sampled around large values of $L$ is too low to allow us to reconstruct an accurate estimation of the posterior in that region. It turns out that by allowing larger chains and varying the covariance matrix steps, the $\chi^2$ landscape is basically almost flat, or at the most with a very small curvature, from very low to large values of $L$, and the MCMC sampling reduces a lot exactly at the top limit, below any possibility to extract a reasonable statistics. 

Thus, in order to try to understand a bit more about it, we used a brute-force filtering mechanism by selecting only those points lying within the $1\sigma$ range with respect to the minimum of the $\chi^2$ ($\Delta \chi^2 =1$ for each single parameter), and then checking the range covered by all the parameters within this range. These are actually the numbers reported in Table~\ref{tab:results} for all parameters.

Reviewing the tables, we can easily see how this approach does not really affect the distribution of the NFW parameters, $c_{200}$ and $M_{200}$, but turns out to be enlightening with respect to the NMC parameter $L$: now, we can conservatively and safely set the range of large $L$ values (corresponding to $L \sim r_s$ as in \citetalias{Zamani:2024oep}) just as \textit{upper limits} to $L$. Although from this result one might equivalently state that very small $L$ values are statistically equivalent to $L \sim r_s$ ones, being these latter values \textit{the largest ones compatible with the data}, actually the very low-rate sampling of the MCMCs around the large values of $L$ is telling us that such region is \textit{more hardly} compatible with the data.

The question to answer now is: is there any \textit{physically strong} reason behind this behaviour? We can get more insights into that by focusing on what our results imply in terms of the mass density and how this reflects in the data which are, after all, the main tool to address the degree of reliability of a model. We have decided to focus on the cluster MACSJ0717, which is not only the only one to exhibit a clear and well-defined constraint on $L$, but this constrain is also at large $L \sim r_s$ (although we note how the value of the Bayesian Factor for it is the largest among all the clusters, so being more disfavored with respect to the GR case).  

\begin{figure}
\centering
\includegraphics[width=8.2cm]{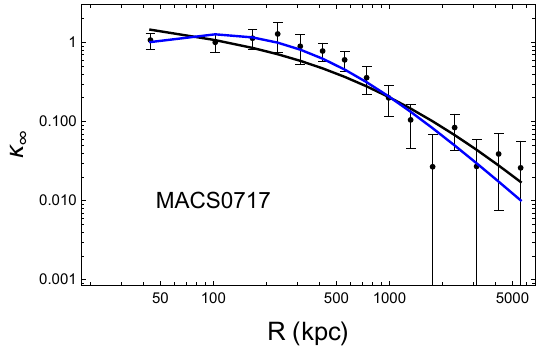}\\
~~~\\
\includegraphics[width=8.2cm]{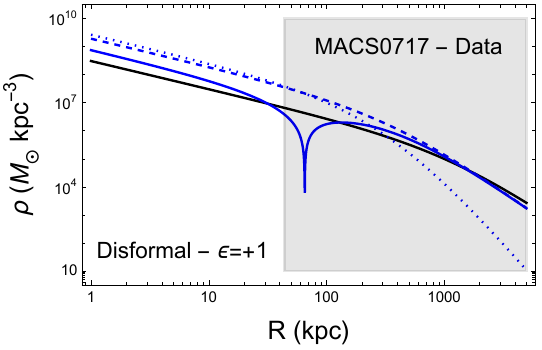}
\caption{\textit{Top:} convergence profiles for the cluster MACSJ0717. The black points (bars) are data (errors) from \textit{CLASH}. The solid blue line corresponds to the disformal scenario with $\epsilon = +1$. 
%The dashed blue line to the disformal case with $(\epsilon = -1)$.
GR is shown as solid black line.
\textit{Bottom:} density profiles for MACSJ0717. Black: $\rho_{\rm DM} + \rho_{\rm gas}$ from GR. Dashed lines: $\rho_{\rm DM} + \rho_{\rm gas}$ from NMC analysis. Dotted line: correction term from the general NMC models, $L^2|\epsilon  \nabla_r \rho_{\rm DM}|$ for the disformal case evaluated at the best fit (corresponding to large values of $L$). Solid line: total contribution, $\rho_{\rm DM} + \rho_{\rm gas}$ plus the corrective term. The grey region shows where we have data.} 
\label{fig: Dens}
\end{figure}

Its convergence profile, both from \textit{CLASH} data and from theory, is shown in the top panel of Fig.~\ref{fig: Dens}. All clusters have data starting from $30\div40$ kpc, but MACS0717 is the only one that exhibits a sort of plateau at small radii and even a slight decrease for $r\lesssim 200$ kpc. Actually, this trend seems to be crucial to explain the large values of $L$ in the NMC scenario when $\epsilon=+1$. There is only one another cluster with a similar profile, MACS0416, although it is much less pronounced.

In the bottom panel of Fig.~\ref{fig: Dens} we dissect the different contributions to Eq.~(\ref{eq: kappamod_Dis}). In black we show the baseline model, the sum $\rho_{\rm DM} + \rho_{\rm gas}$ using the best fits from GR. Dashed lines are the sum $\rho_{\rm DM} + \rho_{\rm gas}$ using the best fits from our NMC scenarios. Dotted lines are the correction terms, $L^2|\epsilon  \nabla_r \rho_{\rm DM}|$ for the disformal case, always evaluated at the best fit (so, with large $L$). Finally, the total contribution, i.e. $\rho_{\rm DM} + \rho_{\rm gas}$ plus the corrective terms, are shown in solid blue. The grey regions are where we have observational data.

We can clearly see how at $\sim 30 \div 40$ kpc, the total contribution from the Poisson equation to the convergence becomes negative, i.e. the corrections from the NMC models are larger than the standard sum $\rho_{\rm DM} + \rho_{\rm gas}$. This happens, of course, for large (enough) values of $L$. As soon as $L$ is small enough to make the correction sub-dominant, the MCMCs tend to prefer low values of $L$, thus making the NMC indistinguishable from GR. That is the reason why assuming our constraints on $L$ as upper limit is the most conservative choice: the $L\sim r_s$ values can be seen as the largest values compatible with data, e.g. the largest values for which the negative contributions from the correction terms do not alter the global profile. 

It is instructive and informative to note that the best constraint for $L$ in the upper range of values happens exactly for MACSJ0717: the trend in convergence at $r<200$ kpc, which seems to slightly decrease, might be explained by a larger negative contribution from the correction terms. In all other objects, the convergence at smaller radii does not decrease at all, and this strongly disfavors very large value of $L$. 

The only reason for that large values are still compatible with the data, even if only as upper limits, is probably due to the errors on the convergence, which leave enough space for the NMC model to fit data with those values. But it is reasonable to think that if we had more precise data at lower distances, we might probably discard definitely the large values of $L$ or, at least, lower those limits from the estimations we have now.   

Furthermore, we can note how the $\epsilon = -1$ case, which actually produces an \textit{always positive} correction to the Poisson equation and to the convergence with respect to GR, although still making possible to have large values of $L$ as upper limits, generally tends to prefer lower values of this parameter which limit the contribution of the correction terms which might spoil the general behavior of the convergence.

\subsection{Single fits with other mass profiles}

{\renewcommand{\tabcolsep}{1.mm}
{\renewcommand{\arraystretch}{2.}
\begin{table*}[ht]
\begin{minipage}{\textwidth}
\centering
\caption{We present stacking results from the analysis of all 19 galaxy clusters using various dark matter profiles. We avoided presenting the upper and lower limit for the parameter $L$ as their large values do not give us useful information.}
\label{tab: Profiles_Stacking}
\scriptsize
\resizebox*{0.95\textwidth}{!}{
\begin{tabular}{c|ccc|cccc|cccc}
\hline
 & \multicolumn{3}{c|}{GR} & \multicolumn{4}{c|}{$\epsilon = -1$} & \multicolumn{4}{c}{$\epsilon = +1$} \\
 Cluster 
 & $c_{200}$ & $M_{200}$ & $\gamma$ 
 & $c_{200}$ & $M_{200}$ & $\gamma$ & $\log L$ 
 & $c_{200}$ & $M_{200}$ & $\gamma$ &  $\log L$ \\
 &           & $(10^{15}\,\mathrm{M}_{\odot})$ & 
 &           & $(10^{15}\,\mathrm{M}_{\odot})$ & &
 &           & $(10^{15}\,\mathrm{M}_{\odot})$ & &
 \\
\hline
\hline
NFW &
$2.64^{+0.22}_{-0.20}$ & $2.38^{+0.33}_{-0.30}$ & $-$ & $2.64^{+0.22}_{-0.20}$ & $2.48^{+0.35}_{-0.31}$ & $-$ & $<0.26$ & $5.15^{+0.80}_{-0.83}$ & $3.09^{+0.48}_{-0.46}$ & $-$ & $3.68^{+0.05}_{-0.05}$ \\
Hernquist &
$5.24^{+0.43}_{-0.39}$ & $0.24^{+0.05}_{-0.04}$ & $-$& $5.24^{+0.43}_{-0.40}$ & $0.24^{+0.05}_{-0.05}$ & $-$ & $<2.19$ & $5.23^{+0.46}_{-0.41}$ &  $0.24^{+0.05}_{-0.05}$ & $-$ & $<3.27$\\
Burkert &
$1.74^{+0.14}_{-0.13}$ & $2.28^{+0.23}_{-0.22}$ & $-$& $1.74^{+0.14}_{-0.13}$ & $2.28^{+0.24}_{-0.22}$ & $-$ & $<3.32$ & $1.74^{+0.14}_{-0.13}$ & $2.28^{+0.22}_{-0.21}$ & $-$ & $<3.33$\\
gNFW &
$3.60^{+0.34}_{-0.36}$ & $2.09^{+0.32}_{-0.28}$ & $0.44^{+0.20}_{-0.21}$ & $3.59^{+0.35}_{-0.35}$ & $2.11^{+0.31}_{-0.28}$ & $0.44^{+0.19}_{-0.21}$ & $<1.79$ & $3.60^{+0.35}_{-0.35}$ & $2.09^{+0.31}_{-0.28}$ & $0.44^{+0.20}_{-0.21}$ & $<2.85$\\
Einasto &
$2.99^{+0.31}_{-0.32}$ & $2.28^{+0.34}_{-0.32}$ & $0.31^{+0.04}_{-0.04}$ & $3.18^{+0.98}_{-0.41}$ & $2.24^{+0.32}_{-0.30}$ & $0.27^{+0.06}_{-0.16}$ & $<4.16$ & $2.98^{+0.30}_{-0.31}$ & $2.30^{+0.35}_{-0.33}$ & $0.31^{+0.04}_{-0.04}$ & $<-0.01$\\
Dark-EXP &
$3.36^{+0.35}_{-0.33}$ & $9.09^{+2.54}_{-1.99}$ & $0.73^{+0.13}_{-0.16}$ & $3.36^{+0.34}_{-0.33}$ & $9.10^{+2.56}_{-2.01}$ & $0.74^{+0.14}_{-0.15}$ & $<3.07$ & $3.37^{+0.34}_{-0.35}$ & $9.01^{+2.57}_{-1.95}$ & $0.74^{+0.14}_{-0.16}$ & $<3.51$ \\
cNFW &
$2.97^{+0.46}_{-0.34}$ & $2.28^{+0.33}_{-0.30}$ & $<164$ & $3.21^{+0.40}_{-0.35}$ & $2.22^{+0.32}_{-0.29}$ & $<36$ & $<2.82$ & $2.66^{+0.22}_{-0.20}$ & $2.38^{+0.32}_{-0.30}$ & $735^{+161}_{-192}$ & $<-33$\\
GPI &
$4.83^{+0.43}_{-0.52}$ & $1.93^{+0.26}_{-0.23}$ & $1.11^{+0.05}_{-0.04}$ & $4.85^{+0.44}_{-0.51}$ & $1.93^{+0.26}_{-0.24}$ & $1.12^{+0.05}_{-0.04}$ & $<-24$ & $4.84^{+0.44}_{-0.51}$ & $1.93^{+0.26}_{-0.24}$ & $1.12^{+0.04}_{-0.04}$ & $<3.01$\\
\hline
\end{tabular}}
\end{minipage}
\end{table*}}}

\begin{figure}
\centering
\includegraphics[width=8.2cm]{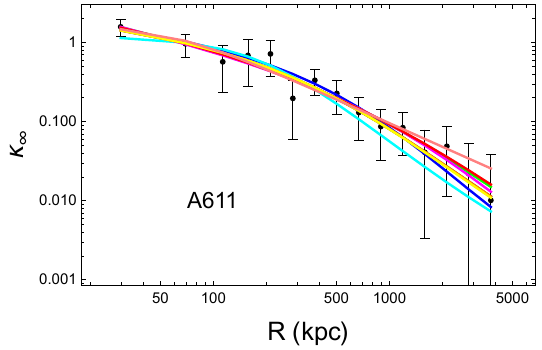}\\
~~~\\
\includegraphics[width=8.2cm]{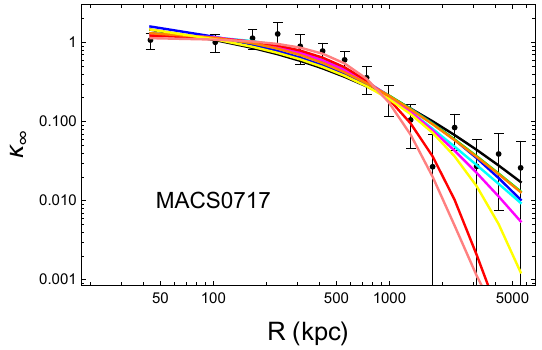}
\caption{Convergence profiles for the cluster A611 (\textit{top panel}) and MACSJ0717  (\textit{bottom panel}). The black points (bars) are data (errors) from \textit{CLASH}. Models: NFW - black; Hernquist - blue; Burkert - cyan; gNFW - green; DARK-exp-$gamma$ - magenta; Einasto - red; cNFW - orange; bNFW - yellow; GPI - pink. All models are from the GR analysis.} 
\label{fig: Models}
\end{figure}

As explained in the previous sections, we have tried to investigate if the found behaviour is related to the specific DM profile we have chosen, the NFW. Thus, we have performed the same analysis, in GR and in both NMC scenarios, using all the other profiles we list in Sec.~\ref{sec: DM} and in \ref{sec: appendix}. In order to get also a visualization of how each and all of them perform with respect to the data, we have decided to plot two representative clusters in Fig.~\ref{fig: Models}: A611 (top panel) is the cluster for which the NFW performs better; MACS0717 (bottom panel) is the object for which performs the worst\footnote{Considering that all clusters have $15$ data points, this selection in terms of best $\chi^2$ is meaningful.}.

For A611, we can see how the profiles do not differentiate too much among them, at least visually speaking. This is confirmed also by the values of the minimum in the $\chi^2$ which is quite similar for all profiles, ranging from $3$ to $4$, and being substantially worse only for the Burkert profile, for which it is $>6.$

For MACS0717, instead, all profiles perform better than NFW, which has a quite large $\chi^2 = 18$, with the best ones being the GPI and the Einasto profile, with $\chi^2<7$. It is interesting to note that these two profiles perform better because they greatly improve the fit in the radii range $[100,1000]\, kpc$, while performing much worse than other models in the outer skirts, at $R>2000$ $kpc$. Apparently, this worst behavior at large $R$ is fully compensated by the improvement at lower $R$.

In general, after looking at all cluster, we can say that the NFW profile is \textit{never} the statistically preferred profile, with the Einasto, DARK-exp-$\gamma$, gNFW and GPI being the ones which perform better. But, unfortunately, even when considering more models, we do not get different results from what we have with NFW: when each cluster is analyzed individually, we obtain very small $L$ and only an upper limit can be set on this parameter.

\subsection{Joint analysis with NFW: universal \texorpdfstring{$L$}{L}}

To achieve a more comprehensive understanding, we conduct also a joint analysis by fitting a universal interaction length ($L$) across all the $19$ clusters, with the hope that an improved constraint on this parameter could be achieved. The results of this joint analysis using a NFW profile, are 
shown in the right half of Table~\ref{tab:results}. They look particularly interesting, as two distinct values for $L$ were obtained depending on the sign of the coupling parameter, $\epsilon$.

For $\epsilon = -1$, a very small value of $L$ was preferred, consistent with the results from the single-cluster fits, in particular when considering the full range of possible values for $L$.

Conversely, for positive $\epsilon$, which introduces a subtractive contribution to the Poisson equation, a larger value for $L$ is observed and quite well constrained. If we compare this new value with the $r_s$ shown in Table 2 of \citetalias{Zamani:2024oep}, we can see that the joint fit approach with $\epsilon = +1$ yields a value for $L$ closer to $r_s$, although substantially smaller (almost one order of magnitude) than our first estimation provided by the Profile Distribution analysis in \citetalias{Zamani:2024oep}.

Although we do not present a table for the other profiles, we conduct a similar joint analysis using the DM profiles described above. Among them, when $\epsilon = +1$, only the Einasto and the cNFW model seem to prefer a (very) negative value of $\log L$, while all other are compatible with, at least, a larger and positive upper limit, thus not moving away from the single-fit analysis, ultimately. In the case of $\epsilon = -1$, only the DARK-exp-$\gamma$, GPI and gNFW models seem to have a preference for an upper limit to $L$, while all other models point to very small values for it.

\subsection{Stacked profiles analysis}

\begin{figure*}
\centering
\includegraphics[width=5.5cm]{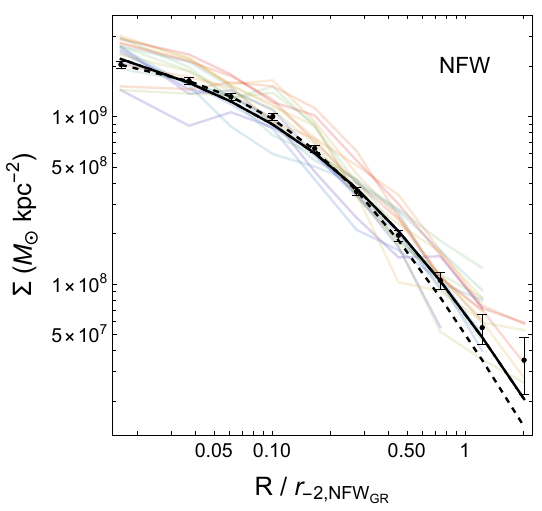}~~~
\includegraphics[width=5.5cm]{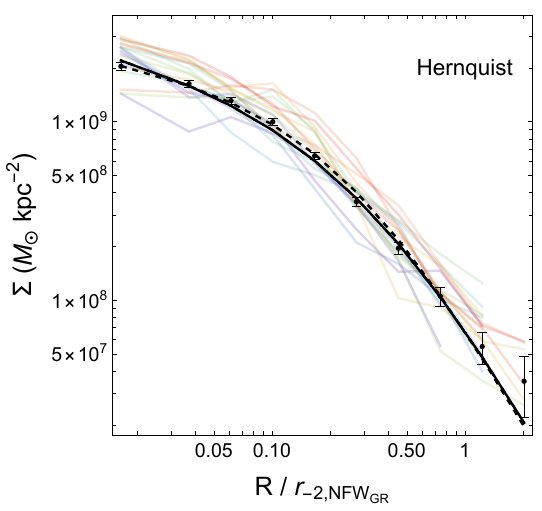}~~~
\includegraphics[width=5.5cm]{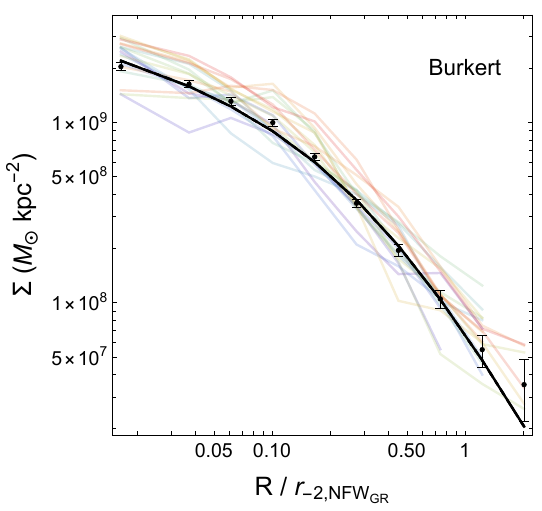}\\
~~~\\
\includegraphics[width=5.5cm]{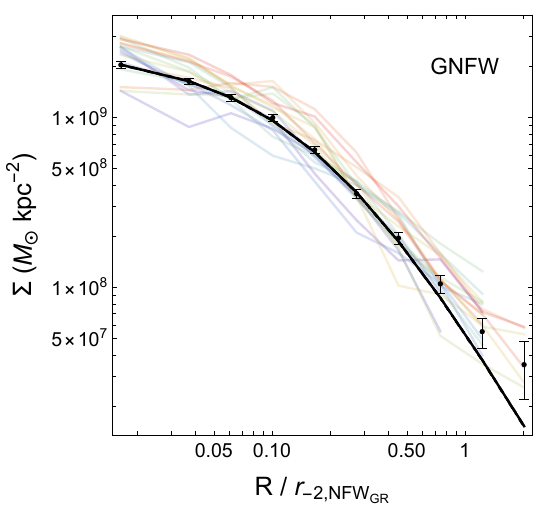}~~~
\includegraphics[width=5.5cm]{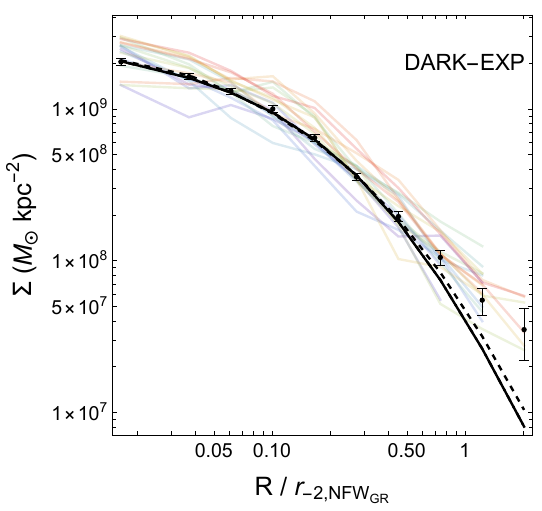}~~~
\includegraphics[width=5.5cm]
{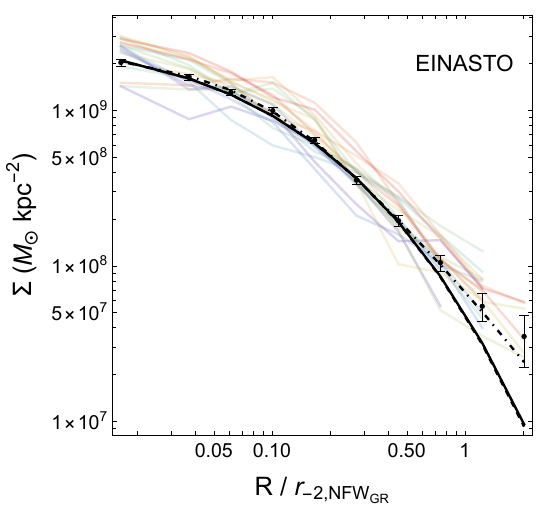}\\
~~~\\
\includegraphics[width=5.5cm]{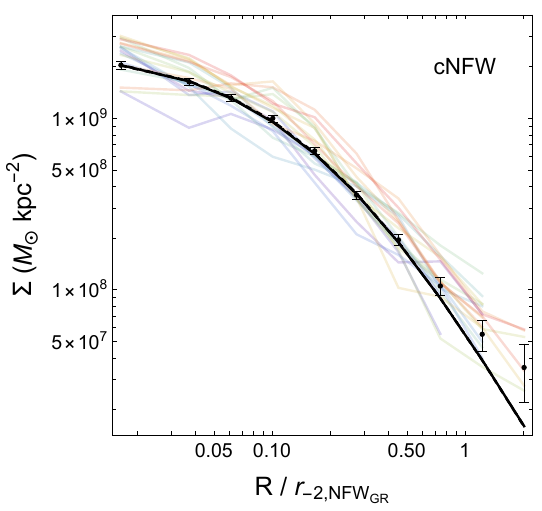}~~~
\includegraphics[width=5.5cm]{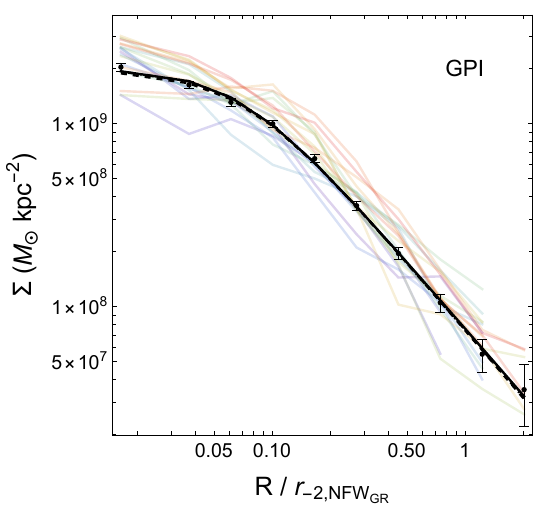}
\caption{Stacked surface density profiles. Black points with error bars: stacked (averaged) densities at the new normalized radius basis (the chosen normalization factor is the value of $r_{-2}$ obtained within GR with a NFW profile, but its value is not decisive for the analysis). Solid black lines: GR; dashed black lines: disformal coupling with $\epsilon=+1$; dot-dashed black lines: disformal coupling with $\epsilon = -1$. Colored lines: profiles from single clusters ordered by redshift from lower $z$ (blue) to higher $z$ (red).} 
\label{fig: Stacking}
\end{figure*}

Moving to the analysis of the stacked profile, a summary of the findings is provided in Table \ref{tab: Profiles_Stacking}. Additionally, the plots in Fig.~\ref{fig: Stacking} show the stacked surface mass density profiles from our sample of galaxy clusters for different mass profiles. Furthermore, the plots feature a set of colored lines representing the individual mass density profiles of the clusters prior to stacking, ordered by their redshift (as they are listed in the Table~\ref{tab:results}). By comparing the colored individual profiles to the averaged stacked profile, one can see how stacking helps to suppress individual peculiarities, thereby revealing a smoother and more universal trend in the data. 

As shown in Table~\ref{tab: Profiles_Stacking}, not even in this case we really get different or deeper insights in the problem. Once again, we are able to fix only an upper limit on the length $L$: it is very small only for the cNFW with $\epsilon=+1$ and GPI with $\epsilon=-1$ cases, while only for the NFW profile with $\epsilon = +1$ the analysis yields a positive and well constrained value for $L$, thus confirming results from the previous sections.

\section{Conclusions}
\label{sec: Conc}

In this study, we have considered Dark Matter to be non-minimally coupled to gravity. Within this scenario, and moving to the weak field limit, one obtains Eq.~\eqref{eq: poissonmoddisf}, a Poisson equation modified by the additional term $\epsilon L^2 \nabla_r \rho_{DM}$. Our main goal was to further explore the case of a disformal coupling, considering both the positive and negative polarities of the parameter $\epsilon$, in order to revise and complete the analysis we have started in \citetalias{Zamani:2024oep}. 

Our findings confirm that, in all scenarios investigated, the Dark Matter parameters $c_{200}$ and $M_{200}$ remain consistent with GR. More importantly, based on the findings in \citetalias{Zamani:2024oep}, where the characteristic interaction length $L$ was found $\propto r_s$ (with $r_s$ the characteristic scale radius of the NFW profile), we wanted to clarify here whether this result was a statistical artifact or represented a consistent physical feature of the model. 

To precisely address this question, we have employed three different approaches to evaluate the strength of our findings: single-cluster fits, joint fit assuming a universal value for $L$, and a stacked analysis of all clusters. Each method was designed to investigate the behavior of the interaction length $L$ under different assumptions and data aggregation strategies, thus providing a comprehensive test of our NMC model.

When looking at individual galaxy clusters, most of them showed very little evidence of any significant differences caused by NMC models. At most, we could set an upper limit $L \sim r_s$. The only exception was the cluster MACSJ0717, where higher well-constrained values of the interaction length $L$ were found. However, we were able to link them to specific features in the convergence profile of this cluster at smaller radii.

In addition to using the widely accepted NFW profile, we also explored other models for dark matter density, such as the Hernquist, Burkert, generalized NFW, cored NFW, DARK-exp-$\gamma$, Einasto, and generalized pseudo isothermal profiles. The aim for analyzing multiple dark matter profiles was to verify if with any of them we could get better constraints on the parameter $L$, also considering that the NFW profile is based on GR simulations, and here we are considering a model alternative to GR. Although some of these models matched the data better than the NFW one, for what concerns the results for the parameter 
$L$, they stayed largely the same. Ultimately, this consistency shows that the findings are not tied to the choice of a specific model.

Then we moved to a joint analysis of all 19 clusters assuming a universal interaction length $L$. The results revealed different behaviors depending on the sign of the coupling parameter $\epsilon$. For $\epsilon = -1$ using NFW profile,  $L$ was consistently constrained to very small values, aligning closely with single-cluster fit results. For $\epsilon = +1$, larger values of $L$ were observed, although these were still smaller than what we got in the \citetalias{Zamani:2024oep}. Analyses with alternative dark matter profiles confirmed these trends, showing either consistent small values or upper limits for $L$, supporting that the observed behaviors do not significantly depart from those identified in the single-cluster fits. 

In addition to the above methods we also did the stacking analysis which provides a clearer and more universal view of the surface mass density profiles across different dark matter models, smoothing out individual cluster variations. While this method confirmed general trends, it did not reveal significant new insights.
In most cases, the interaction length $L$ was constrained to small values or treated as an upper limit. Notably, only the NFW profile with $\epsilon=+1$ produced a well-constrained positive value for $\log L$, while in all other profiles it generally gives us an upper limit. Overall, the stacking analysis supports earlier results, showing that $L$ has minimal impact in most scenarios. 

Thus, the main conclusion we can infer is that very likely clusters of galaxies are not the best probe to test this model, probably because of their scale, and/or because the specific observational probe we have used here (gravitational lensing) is not sensitive (large errors) enough to the modifications introduced by this model. We have planned to test this theoretical scenario on smaller astrophysical scales, like with kinematical and dynamical data from galaxies, in order to collect further information about its reliability as contender to the standard dark matter paradigm.

\section{acknowledgements}
This manuscript has no associated data
or the data will not be deposited. [Authors’ comment: No new data
have been produced during the study. All used data are available upon
reasonable request to/from Keiichi Umetsu.]

The research of S.Z. and V.S. is funded by the Polish National Science Centre grant No. DEC-2021/43/O/ST9/00664. D.B. acknowledges support from  projects PID2021-122938NB-I00 funded by the Spanish “Ministerio de Ciencia e Innovación” and FEDER “A way of making Europe”, PID2022-139841NB-I00
funded by the Spanish “Ministerio de Ciencia e Innovación” and SA097P24 funded by Junta de Castilla y León.

\appendix
\section{Dark matter Halo Profiles}
\label{sec: appendix}

Here we briefly review the halo profiles that we have considered in our analysis. 
%%%%%%%%%%%%%%%%%%%%%%%%%%%%%%%%%

For the Hernquist profile \cite{Hernquist:1990be} we have
\begin{align}
   \rho(r) & = \frac{\rho_s}{\left(\frac{r}{r_s}\right) \left(1 + \frac{r} {r_s}\right)^3}\; , \\
   r_{-2} &= \frac{r_s}{2}\, , \nonumber \\
   \rho_{s} &= \frac{{\rm \Delta}}{3} \rho_{c}\, c_{{\rm \Delta}} \left(1+\frac{c_{{\rm \Delta}}}{2}\right)^{2}\, . \nonumber
\end{align}
%%%%%%%%%%%%%%%%%%%%%%%%%%%%%%%%%

The Burkert Profile is \cite{Burkert:2000di, Mori:2000me}
\begin{align}
    \rho(r)&=\frac{\rho_s}{\left(1+\frac{r}{r_s}\right)\left(1+\frac{r^2}{r_s^2}\right)}\;, \\
    r_{-2}& \approx 1.521 \cdot r_s\, , \nonumber \\
    \rho_s& = \frac{{\rm \Delta}}{3} \frac{\rho_{c} \left(1.521 c_{{\rm \Delta}}\right)^{3}}{-\frac{\arctan \left(1.521 c_{{\rm \Delta}}\right)}{2} 
    + \frac{\log \left[(1+(1.521 c_{{\rm \Delta}})^2)(1+1.521 c_{{\rm \Delta}})^2\right]}{4}} . \nonumber
\end{align}

%%%%%%%%%%%%%%%%%%%%%%%%%%%%%%%%%

The density profile for the cored NFW (cNFW) model is given as \cite{Newman:2012nv,Newman:2012nw}
\begin{align}
\rho(r) &=
\frac{\gamma \rho_s} {\left(1 + \frac{\gamma r}{r_s}\right) \left(1 + \frac{r}{r_s}\right)^{2}},\\
r_{-2} &= r_s \left( \frac{\tilde{\gamma}}{2 \gamma}\right)\, , \nonumber \\
\rho_{s} &= \frac{{\rm \Delta}}{3}\rho_{c} c_{{\rm \Delta}}^3
\frac{(\gamma -1)^2 \tilde{\gamma}^3}{8 \gamma ^4 \left(\frac{(1-\gamma ) \tilde{\gamma} c_{{\rm \Delta}}}{2 \gamma  \left(\frac{\tilde{\gamma} c_{{\rm \Delta}}}{2 \gamma }+1\right)}+\log
   \frac{\left(\frac{1}{2} \tilde{\gamma} c_{{\rm \Delta}}+1\right)^{\frac{1}{\gamma}}}
   {\left(\frac{\tilde{\gamma} c_{{\rm \Delta}}}{2 \gamma }+1\right)^{2-\gamma}}\right)} \, , \nonumber
\end{align}
with $\tilde{\gamma} = \gamma + \sqrt{\gamma \left( 8 + \gamma \right)}$ and $\gamma>1$.

The generalized NFW (gNFW) profile \cite{Umetsu:2015baa,Zhao:1995cp} is defined by
\begin{align}
    \rho(r)&=\frac{\rho_s}{\big(\frac{r}{r_s}\big)^{\gamma} \big(1+\frac{r}{r_s}\big)^{3-\gamma}}\;, \\
    r_{-2}&= (2-\gamma) r_s\;, \nonumber \\
    \rho_s &=\frac{{\rm \Delta}}{3}\rho_{c} c_{{\rm \Delta}}^{\gamma} \frac{(3-\gamma) (2-\gamma)^{\gamma}}{ 
 _{2}F_{1}[3-\gamma,3-\gamma,4-\gamma,(\gamma-2) c_{{\rm \Delta}}]
}\, , \nonumber
\end{align}
where $_{2}F_{1}$ is the Gaussian hypergeometric function, and $0<\gamma<2$.

For the DARK-exp-$\gamma$ profile we have \cite{Umetsu:2015baa,Hjorth:2015bfa}
\begin{equation}  
\begin{aligned}
        \rho(r) &= \rho_s \left(\frac{r}{r_s}\right)^{-\gamma} \left(1+\frac{r}{r_s}\right)^{\gamma-4}\; ,\\ 
        r_{-2}&= \left(1-\frac{\gamma}{2}\right)r_s\, , \nonumber \\
        \rho_s &= \frac{{\rm \Delta}}{3}\rho_{c} c_{{\rm \Delta}}^3 \left(1-\frac{\gamma}{2}\right)^3 \frac{3-\gamma}{\frac{(1-\gamma/2) c_{{\rm \Delta}}}{1 + (1-\gamma/2) c_{{\rm \Delta}})}}\, , \nonumber
\end{aligned}
\end{equation}
with $0<\gamma<2$.

The Einasto profile \cite{Einasto:1965czb,Retana-Montenegro:2012dbd} is defined by
\begin{align}
    \rho(r) &= \rho_s \exp\left\{-\frac{2}{\gamma} \left[\left(\frac{r}{r_s}\right)^{\gamma} - 1\right]\right\},\\
    r_{-2}&= r_s\, , \nonumber \\
%    \rho_s &= \frac{{\rm \Delta}}{3} \rho_{c} c_{{\rm \Delta}}^3 \frac{\gamma \exp\left[-\frac{2}{\gamma}\right]}{\frac{\gamma}{2}^{\frac{3}{\gamma}}} \frac{1}{{\rm \Gamma}\left(\frac{3}{\gamma}\right)-{\rm \Gamma}\left(\frac{3}{\gamma},\frac{c_{{\rm \Delta}}^{\gamma}}{\frac{\gamma}{2}}\right)}\, , \nonumber \\
    \rho_s &= \frac{{\rm \Delta}}{3} \rho_{c} c_{{\rm \Delta}}^3 \frac{\gamma \exp\left[-\frac{2}{\gamma}\right] \left(\frac{2}{\gamma}\right)^{\frac{3}{\gamma}}}{{\rm \Gamma}\left(\frac{3}{\gamma}\right)-{\rm \Gamma}\left(\frac{3}{\gamma},\frac{2}{\gamma}c_{{\rm \Delta}}^{\gamma}\right)}\, , \nonumber
\end{align}
where ${\rm \Gamma}(z)$ and ${\rm \Gamma}(a,z)$ are respectively the Euler and incomplete gamma functions, and $0<\gamma<2$.

Finally, we have defined the Generalized Pseudo Isothermal (GPI) as
\begin{align}
\rho(r) & = \frac{\rho_s}{\left[1 + \left(\frac{r}{r_s}\right)^2\right]^\gamma}\, , \\
r_{-2} & = \frac{r_s}{\sqrt{\gamma-1}}\, , \nonumber \\
\rho_s &= \frac{{\rm \Delta}}{3} \rho_{c}
\frac{\frac{c^{4}_{{\rm \Delta}}}{\left(\gamma-1\right)^2}}
{\left[ 1 + \frac{c^{2}_{{\rm \Delta}}}{\gamma-1} \right]^{\gamma-1}} \nonumber \\
&\cdot \frac{3+4(\gamma-2)\gamma}
{-1+\frac{(1-2\gamma)}{\gamma-1}c^{2}_{{\rm \Delta}}+{_2}F_{1} \left[1,\frac{1-2\gamma}{2},\frac{1}{2},-\frac{c^{2}_{{\rm \Delta}}}{\gamma-1}\right]}\, , \nonumber
\end{align}
with $\gamma>1$.

\bibliographystyle{apsrev4-1}
\bibliography{bibliography.bib}

\end{document}